\documentclass{aa}

\usepackage{graphicx}
\usepackage{natbib,twoopt}
\usepackage[breaklinks=true]{hyperref} 
 \bibpunct{(}{)}{;}{a}{}{,}    
 \newcommandtwoopt{\citeads}[3][][]{\href{http://adsabs.harvard.edu/abs/#3}%
                                        {\citealp[#1][#2]{#3}}}
 \newcommandtwoopt{\citepads}[3][][]{\href{http://adsabs.harvard.edu/abs/#3}%
                                        {\citep[#1][#2]{#3}}}
 \newcommandtwoopt{\citetads}[3][][]{\href{http://adsabs.harvard.edu/abs/#3}%
                                        {\citet[#1][#2]{#3}}}
 \newcommandtwoopt{\citeyearads}[3][][]%
   {\href{http://adsabs.harvard.edu/abs/#3}{\citeyear[#1][#2]{#3}}}
\bibpunct{(}{)}{;}{a}{}{,}
\usepackage{latexsym}
\usepackage{subfigure}
\usepackage{amsmath}
\usepackage{amssymb}
\usepackage{amstext}
\usepackage{color}
\usepackage{psfrag}
\usepackage{txfonts}

\def\cm3{cm$^{-3}$}

\def\12{$^{12}$CO}

\begin{document}

\title{
Unveiling the origin of HESS~J1809$-$193}

\author{G. Castelletti\inst{1}\fnmsep
\and 
E. Giacani\inst{1,2}
\and A. Petriella \inst{1,3}
}

\offprints{G. Castelletti}
\institute {Instituto de Astronom\'{\i}a y  F\'{\i}sica del Espacio (IAFE),UBA-CONICET,
CC 67, Suc. 28, 1428 Buenos Aires, Argentina\\
             \email{gcastell@iafe.uba.ar}
\and
FADU, University of Buenos Aires, Buenos Aires, Argentina
\and
CBC, University of Buenos Aires, Buenos Aires, Argentina}

   \date{Received <date>; Accepted <date>}


\abstract 
{}
 {
The main goal of this paper is to provide new insights on the origin  of the observable flux 
of $\gamma$ rays from HESS~J1809$-$193 using new high-quality observations in the radio domain.}
  {We used the Expanded Very Large Array (now known as the Karl G. Jansky Very large Array, JVLA) to produce 
a deep full-synthesis imaging at 1.4~GHz of the vicinity of PSR~J1809$-$1917. These data were used in 
conjunction with $^{12}$CO observations from the James Clerk Maxwell Telescope in the transition 
line J=$3-2$ and atomic hydrogen data from the Southern Galactic Plane Survey to investigate the properties 
of the interstellar medium in the direction of the source HESS~J1809$-$193. 
  }
  {The new radio continuum image, obtained with a synthesized beam of 
8$^{\prime\prime}$$\times$ 4$^{\prime\prime}$ 
and a sensitivity of 0.17~mJy~beam$^{-1}$, reveals with unprecedented detail all the intensity 
structures in the field. No radio counterpart to the observed X-ray emission supposed to be a 
pulsar wind nebula 
powered by PSR~J1809$-$1917 is seen in the new JVLA image. We discovered a system of molecular clouds on 
the edge of the supernova remnant (SNR) G11.0$-$0.0 shock front, which is positionally coincident 
with the brightest part of the TeV source HESS~J1809$-$193. We determine, on the basis of kinematic and morphological evidences, a physical link of the SNR with the clouds for which we estimated a total 
(molecular plus atomic) mass of $\sim$3$\times$$10^{3}$~M$_{\odot}$ and a total proton density in the 
range 2-3$\times$$10^{3}$~cm$^{-3}$. 
}
   {We propose as the most likely origin of the very high-energy $\gamma$-ray radiation from HESS~J1809$-$193 a hadronic mechanism through collisions of ions accelerated at the SNR G11.0$-$0.0 shock  with the molecular matter in the  vicinity of the remnant.} 


\maketitle
\titlerunning{Unveiling the origin of HESS~J1809$-$193}
\authorrunning{Castelletti et al.}

\section{Introduction}
Since their discovery, Galactic  $\gamma$-ray sources have been linked to a variety of 
astrophysical phenomena like X-ray binaries, star-forming material, remnants of supernova
explosions, and pulsar wind nebulae (PWNs), among others 
\citep{denaurois2015}. The importance of a broadband approach using multiwavelength observations for 
understanding the $\gamma$-ray emission processes has been firmly demonstrated over the course of the
last few years. Good examples of this class of studies performed especially in the radio and X-rays bands are those presented by \citet{aharonian2008}, \citet{abramowski2011}, \citet{gotthelf14}, on the $\gamma$-ray 
sources named
\object{HESS~J1713$-$381} (linked to the supernova remnant (SNR) \object{CTB~37B}), 
\object{HESS~J1731$-$347} (in SNR~\object{G353.6$-$0.7}), and 
\object{HESS~J1640$-$465} (in \object{PSR~J1640$-$4631}, SNR~\object{G338.3$-$0.0}), respectively. Surprisingly, there is still a significant fraction of cases for which the production mechanisms of $\gamma$ rays has 
not been established\footnote{For a list of the discovered VHE $\gamma$-ray sources see the online TeV 
$\gamma$-ray catalog TeVCat http://tevcat.uchicago.edu.}.

In this paper we focus on the very high-energy (VHE) source  HESS~J1809$-$193. It was discovered
during the systematic search for VHE $\gamma$ radiation from pulsars in the Galactic plane survey
performed by the High Energy Stereoscopic System (H.E.S.S.) Collaboration \citep{Aharonian07}. 
Since its first detection, new data were recorded increasing the detection significance
considerably to a value well above 10$\sigma$. The detection of the TeV source occurs
in an area of radius $0^{\circ}.25\pm0^{\circ}.02$. Fitting  the excess map with a 2D
symmetric Gaussian, \citet{renaud2008} estimated its best-fit position 
at R.A.=$18^{\mathrm{h}}\,09^{\mathrm{m}}\, 52^{\mathrm{s}}$, 
Dec.=$-19^{\circ}\,23^{\prime}\,42^{\prime\prime}$ (J2000.0).

The peak of the VHE emission lies about 6$^{\prime}$ to the south of \object{PSR~J1809$-$1917}, 
a middle-aged ($\tau$ =51~kyr) and energetic (spin-down luminosity 
$\dot{E}$= 1.8 $\times 10^{36}$~erg~s$^{-1}$) 82.7~ms radio pulsar, discovered during the Parkes 
Multibeam Pulsar Survey of the Galactic Plane \citep{Morris02}. 
Using dispersion measure data, distances of $\sim3.5$ and $\sim3.7$~kpc were derived to this pulsar by 
\citet{Cordes02} and \citet{Morris02}, respectively. 
In the X-ray domain, on the basis of {\em Chandra} observations, \citet{Kargaltsev07} 
reported the detection of emission from PSR~J1809$-$1917 
and synchrotron emission from the associated pulsar wind nebula. 
The overall X-ray morphology of the nebula is dominated by a bright inner component, 
$3^{\prime\prime}\times12^{\prime\prime}$ in diameter, elongated in the north-south 
direction (with the pulsar close to its south end), which in turn is immersed in a halo of lower 
surface brightness $\sim$ $20^{\prime\prime}\times40^{\prime\prime}$ in size and  elongated 
in the same direction. The energy spectrum of the PWN is characterized by a power law index $\Gamma=1.4$, 
while its luminosity is $L\sim4\times10^{32}$~erg~s$^{-1}$ between 0.8 and 7~keV at a distance of 3.5~kpc. 
The X-ray nebula is also embedded in a fainter but still clearly detectable emission that extends $\sim4^{\prime}$ southward of the pulsar \citep{Kargaltsev07}. Later on, a study performed with the {\em Suzaku} 
satellite confirmed  the existence of diffuse non-thermal X-ray emission away from the pulsar up to 
$\sim20^{\prime}$ \citep{Anada10} in the direction of HESS~J1809$-$193. On the basis of this detection along 
with the fact that the pulsar is energetic enough to power the bulk of the observed $\gamma$-ray emission, PSR~J1809$-$1917 and its nebula are considered  the most likely counterpart to HESS~J1809$-$193 to date. 
The offset of the VHE centroid from the pulsar position is not atypical  as is exemplified by 
Vela pulsar \object{PSR~B0833$-$45} associated with HESS~J0835$-$455 \citep{aharonian2006a} and 
\object{PSR~B1823$-$13} with  \object{HESS~J1825$-$137} \citep{aharonian2006b}. 
This PWNe offset can be explained by either a high spatial velocity of the pulsar \citep{vanderSwaluw04} 
or the evolution of the SNR blastwave into an inhomogeneous interstellar medium (ISM). Thus, an asymmetric 
reverse SNR shock reaches one side of the PWN sooner than the other side, crushing the PWN 
\citep{Blondin01}. 
This scenario has been proposed for the PSR~J1809$-$1917/HESS~J1809$-$193 system \citep{Kargaltsev07} and 
could be tested through the detection of radio emission from the relic electrons of the crushed PWN. 

Remarkably, the presence of other possible sites of energetic particles such as radio supernova remnants 
and HII regions within the extension of HESS~J1809$-$193  hampers a decisive conclusion on the origin 
of the $\gamma$ rays and makes it necessary to revise the PWN scenario considered for the VHE source.

In this paper, we report on new high-quality synthesis imaging obtained from observations carried out with 
the Expanded Very Large Array (or the Karl G. Jansky Very Large Array as it is now commonly known) of a large 
region toward HESS~J1809$-$193. The purposes of our research are the detection of the radio counterpart 
to the proposed X-ray PWN powered by PSR~B1809$-$1917 and/or any trace of the alleged host SNR of this pulsar, 
as well as the identification of other potential particle accelerators in the field as alternative candidates 
that produce the VHE $\gamma$-ray emission. Additionally, we outline the discovery of molecular material adjacent 
to the SNR~\object{G11.0$-$0.0} well within the perimeter of HESS~J1809$-$193 and analyze its relationship with the production of the detected $\gamma$-ray photons through a hadronic mechanism.

\section {Data}
\subsection {New radio observations}
The radio continuum observations at L-band were performed under the shared-risk commissioning phase  with the Expanded Very Large Array (EVLA\footnote{Hereafter,  we  use the acronym JVLA 
to refer to either data or imaging products obtained from the Expanded Very Large Array observations carried 
out under our program code 12A-166.}) of the National Radio Astronomy Observatory\footnote{The National Radio Astronomy Observatory (NRAO) is a facility of the National Science Foundation operated by Associated 
Universities, Inc. under cooperative agreement with the National Science Foundation.} in its B and C array configurations as summarized in Table~\ref{evla-data} (program code 12A-166). The phase center of the 
observations was located at R.A.=$18^{\mathrm{h}}\,09^{\mathrm{m}}\,43^{\mathrm{s}}$, 
Dec.=$-19^{\circ}\, 17^{\prime}\, 38^{\prime\prime}$ (J2000.0). 
The correlator was configured to give a 1~GHz wide-band in full polarization. The
baseband was comprised of sixteen 64~MHz sub-bands or spectral windows (SPWs) containing
64 frequency channels. Both in B- and C-array data, significant interference was found especially affecting 
the SPWs in the frequency range $\sim$1.5-1.7~MHz, which was entirely removed.
In addition, ten channels at the edges of the SPWs were flagged owing to the roll-off of
the digital filters in the signal chain. Our target was observed at intervals of ten minutes, each bracketed 
by the phase calibrator source \object{J1834$-$1237}. The standard calibrator \object{3C~286} 
was used for flux and bandpass. The flux density scale was set according to the Perley-Butler coefficients derived at the JVLA in 2010. 

\begin{table}
\caption{Summary of JVLA observations}
\renewcommand{\arraystretch}{1.0}                                                                                
\begin{center}
\begin{tabular}{ccc} \hline\hline \\
Configuration & Date         &   Hours \\
\hline
C             &    2012 Feb 5 &  1.5  \\
C             &    2012 Feb 18&  1.5   \\
B             &    2012 Jun 20& 1.5   \\
B             &    2012 Jun 21 & 1.0   \\  \hline
\label{evla-data}
\end{tabular}
\end{center}
\end{table}

Data processing for each observing run of the JVLA was fully performed with the Common Astronomy Software Applications (CASA) package. For each dataset, after updating the antenna positions and removing obviously corrupted data by hand, we determine appropriate complex antenna gains for the calibrators, which were then applied to the target data. The final calibrated visibility data from all the available observations for each configuration were combined into a single \it uv\rm-data set, before starting the imaging process. During deconvolution we employed the Multi-Scale Multi-Frequency Synthesis imaging algorithm \citep{rau11} to simultaneously deal with the multiple scales and the variations of the spectral indices of the sources in the field. In addition, we used the CASA implementation of the W-Projection algorithm to correct for the sky curvature. The imaging process was performed by adopting the Briggs robust parameter equal to 0, a 
compromise between the natural and uniform weighting schemes of visibilities.

\begin{figure*}[!ht]
\centering
\includegraphics[scale=0.7]{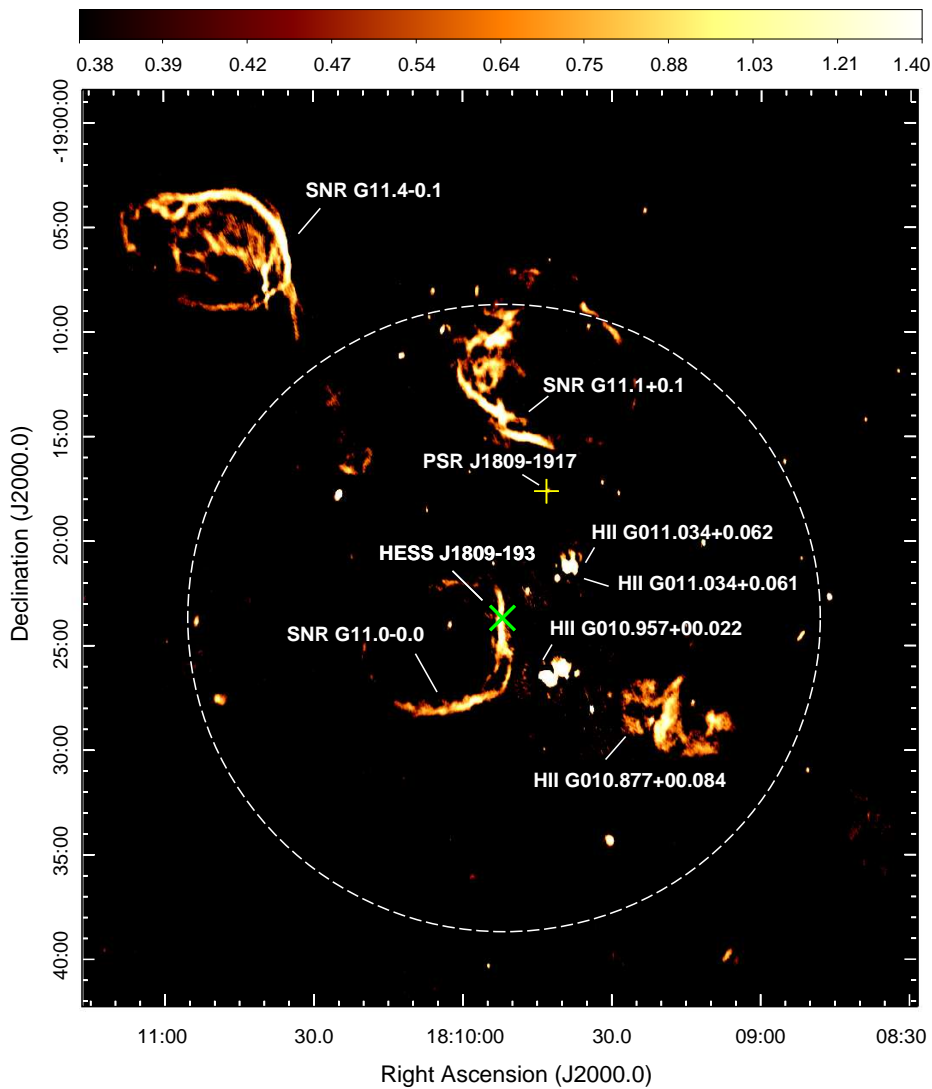}
\caption{Radio continuum image at 1.4~GHz of the region surrounding the pulsar PSR~J1809$-$1917 constructed
using the C and B configurations of the JVLA. The intensity scale is based on a square root relation in units of mJy~beam$^{-1}$. The final beam size is 8$^{\prime\prime}$.3 $\times$ 4$^{\prime\prime}$.4 with a position 
angle of $\sim$$-$170$^{\circ}$. The noise level is 0.17~mJy~beam$^{-1}$ without considering attenuation of the primary beam. The names of SNRs and HII regions lying within the field are indicated. 
The location of PSR~J1809$-$1917 is shown with a yellow plus sign, while the best-fit position of the centroid of HESS~J1809$-$193 is indicated with a green cross. The white dashed circle denotes the 
measured extension of the VHE source in terms of the sigma (rms) of a symmetric 2D Gaussian fit to the TeV data 
\citep{renaud2008}. 
}
\label{FoV-G11}
\end{figure*}

After combining the data from the C and B configurations the resulting FWHM of the synthesized beam is 
8$^{\prime\prime}$.3$\times$4$^{\prime\prime}$.4, P.A.=-169$^{\circ}$.62, and the noise level is 
0.17~mJy~beam$^{-1}$. The final radio map of the field is shown in Fig.~\ref{FoV-G11}. This is the first 
high-resolution image of the region towards HESS~J1809$-$193 that provides unprecedented sensitivity to the 
radio emission surpassing by more than an order of magnitude the sensitivity of available radio images at the same frequency. As can be seen from this figure, this is a complex region of our Galaxy, full of SNRs and HII regions, all of them indicated in the figure.

\subsection { Surrounding medium}
The interstellar medium properties of the region around HESS~J1809$-$193 were investigated using the neutral
hydrogen (HI) data from the Southern Galactic Plane Survey \citep{Mc05} which maps the atomic gas with an angular resolution of 2$^{\prime}$, a velocity resolution of 0.82~km~s$^{-1}$, and an rms sensitivity of 1~K. We also explored the $^{12}$CO~J=$3-2$ transition in the region using observations extracted from the data archive of the James Clerk Maxwell Telescope (JCMT, Mauna Kea, Hawaii). The observations (project ID: M10AU20) were obtained using the HARP-ACSIS instrument and have angular and  spectral resolutions of 14$^{\prime\prime}$ and 0.42~km~s$^{-1}$, respectively. We used the reduced data produced by the standard ORAC-DR pipelines, which are presented in units of  corrected antenna temperature $T_{A}^{*}$. To convert to main beam temperature $T_{mb}$, we applied the relation $T_{mb}=T_{A}^{*}/\eta_{mb}$, where $\eta_{mb}$ is the main beam efficiency. Following \citet{buckle09}, we used $\eta_{mb}=0.6$. 

\section{Results and discussion}
\subsection{ TeV emission from a PWN}
Figure~\ref{FoV-G11} shows the region covered by the PSR~J1809$-$1917/HESS~J1809$-$193 system. 
From the new image no evidence is found, down to the noise level of our data, for the radio 
counterpart to either the X-ray PWN  or the surrounding radio shell associated with its host SNR.

\begin{figure}
\center
\vspace{1cm}
\includegraphics[scale=0.50]{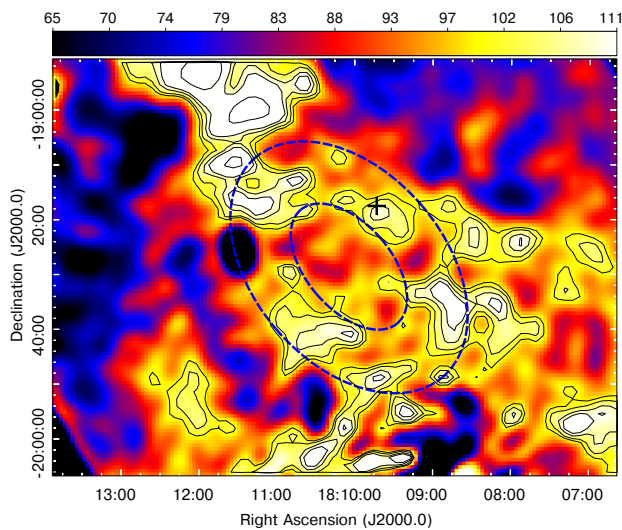}
\caption{Neutral hydrogen distribution towards PSR~J1809$-$1917 and its nebula.
The overlaid contours represent the HI emission at 100, 103, 107, 110, and 120~K. 
The blue dashed annulus approximately delimits the higher density gas of the region, while the 
plus sign marks the position of PSR~J1809$-$1917.}
\label{HI-shell}
\end{figure}

There are many pulsars for which the search of radio PWNe yielded negative results
\citep[e.g.,][]{Gaensler00,Giacani09,Giacani14}. In these cases, the failure to detect the nebular emission 
has been explained as a consequence of two main physical conditions based on either a high magnetic field 
that inhibits the production of synchrotron radiation at longer wavelengths 
or severe adiabatic losses that occur in young and energetic pulsars lying in a very low ambient density 
($\sim$~0.003~cm$^{-3}$). In the case of PSR~J1809$-$1917, using the parameters previously derived for this 
object \citep[$P$=82.7~ms, $\dot{P}$=2.55$\times$10$^{14}$,][]{Morris02}, we obtain a pulsar magnetic field strength $B_{p}$=3.2$\times$10$^{19}\,(P\dot{P})^{1/2}$=1.5$\times$ 10$^{12}$~G 
in good agreement with that derived for most young pulsars in SNRs \citep{Gaensler06}. 
On the other hand, to quantify the properties of the ambient medium where the pulsar is
propagating we inspected the structure of the HI emission in a large region surrounding
PSR~J1809$-$1917 in the velocity interval 28-31~km~s$^{-1}$ 
(all the velocities mentioned in this work are relative to the local standard of rest)
corresponding to the distance range estimate to PSR~J1809$-$1917 of 3.5-3.7~kpc \citep[according to the 
rotation curve model of the Milky Way of][with the galactic distance 
R$_{0}$=8.5~kpc and the rotation velocity at the Sun $\Theta$=220~km~s$^{-1}$]{Fich89}.
Figure~\ref{HI-shell} shows a HI large-scale feature surrounding the pulsar, which at first glance looks like a blown wind bubble generated by the stellar wind of the progenitor star. We crudely estimate the volume density of the ambient atomic gas in the cavity by integrating the column density distribution 
$N_{\mathrm{H}}$ within the mentioned velocity range and assuming for the cavity a spherical geometry with a radius of $\sim$ 9$^{\prime}$ 
at a distance of 3.6~kpc (an average between the 3.5 and 3.7~kpc).
After subtracting an appropriate background level to account for diffuse emission that may come from far gas 
whose emission is detected at the same radial velocities in this direction of the Galaxy, 
the obtained ambient medium density is  $\sim$1.5~cm$^{-3}$. 
Regardless of whether this HI structure represents  the imprint left in the ISM by a past event that ended in a SN explosion or simply inhomogeneities in the medium, the obtained value rules out the 
possibility of an expansion in an area of very low density. 
We consequently note that the conditions invoked to account for the lack of an associated
radio nebula fail to explain the case of PSR~J1809$-$1917.

Nevertheless, we can still speculate that if the radio pulsar wind nebula existed and if
it were at least 4$^{\prime}$ across (that is, a size similar to that of the faintest 
X-ray emission observed with \it Chandra\rm), then the resulting upper limit for the nebular flux density at 1.4~GHz -- consistent with our non-detection (at a sensitivity of 0.2~mJy~beam$^{-1}$) -- would be 
$\sim$250~mJy. 
Assuming a typical PWN radio spectral index $\alpha$ =$-$0.3 \citep[S $\propto \nu^{\alpha}$,][]{Gaensler06}, 
we find that the mentioned flux at 1.4~GHz corresponds to a broadband radio luminosity integrated 
between 10$^{7}$ and 10$^{11}$~Hz of $L_{R} \sim 1.4 \times 10^{32}$~erg~s$^{-1}$ (at a distance of 3.6~kpc) 
and to an efficiency $\epsilon \equiv L_{\mathrm{R}}/ \dot{E}$$\sim 7.7\times 10^{-5}$; both estimates
are significantly lower than the typical values of $L_{\mathrm{R}} \sim 10^{34}$~erg~s$^{-1}$ 
and $\epsilon \ge 10^{-4}$ observed in other radio PWNe \citep{Gaensler06}. 
However, this analysis should be taken with  caution as exceptions exist; for example,
the pulsars \object{PSR~B1706$-$44} and \object{PSR~B0906$-$49} both power 
observable radio nebulae with very low efficiencies of about $10^{-6}$ \citep{Giacani01}. 
 
As  mentioned in Sect.~1, PSR~J1809$-$1917 appears to be the most viable counterpart to the 
HESS~J1809$-$193 source in a leptonic scenario where the reverse shock from a surrounding
SNR has collided with the entire shock surface bounding the nebula powered by PSR~J1809$-$1917.
After this phase an offset nebula is formed and HESS~J1809$-$193 may correspond to such nebula emitting 
in the VHE domain. Numerical simulations of this process estimate a crushing timescale of 
$\leq 10^{3}\,E_{51}^{-1/2}(M_{ej}/M_{\odot})^{5/6}\eta^{-1/3}$~yr, where $ E_{51}$ is the total 
mechanical energy of the SN explosion in units of 10$^{51}$~erg, $M_{ej}$ is the ejected mass, 
and $\eta$ is the ambient hydrogen density in cm$^{-3}$ \citep{vanderSwaluw04}.
In the case of PSR~J1809$-$1917, if we assume that $E_{51}$ is equal to unity, $M_{ej}$=10~$M_{\odot}$ and 
we use the hydrogen density derived from our study of the HI distribution in the region of the pulsar 
($\eta$=1.5~cm$^{-3}$), we obtain a crushing timescale of $\sim$6~kyr. This time is considerably lower than 
the characteristic age of the pulsar even if we consider an initial energy of the explosion of 10$^{50}$~erg.
This relic PWN scenario requires the existence of a SNR, but taking into consideration the 
characteristic age of PSR~J1809$-$1917, the radio emission of the alleged remnant 
might have dissipated through the action of radiative shocks, making the SNR forward shock undetectable.
Thus, a relic TeV PWN origin for HESS~J1809$-$193 cannot be dismissed. It should be pointed, however,  out that this scenario is not completely consistent with the location of the pulsar in the 
cometary shaped nebula observed in X-rays in the vicinity of PSR~J1809$-$1917. Indeed,
as noticed by \citet{Anada10} both \it Chandra \rm and \it Suzaku \rm observations suggest 
that the pulsar is moving towards the center of the HESS~J1809$-$193 source. 

\subsection {TeV emission from a hadronic mechanism}
In our search for an alternative explanation of the observed $\gamma$-ray emission, the 
SNR~G11.0$-$0.0 that lies in projection within the brightest region of the VHE source
appears to be a promising candidate for the acceleration of particles via $\pi^{0}$-decay emission
arising from proton-proton collisions. In this context, we search for spatial correspondences 
between matter concentration and the $\gamma$-ray photons observed in HESS~J1809$-$193.

\begin{figure*}[ht!]
\centering
\includegraphics[scale=1]{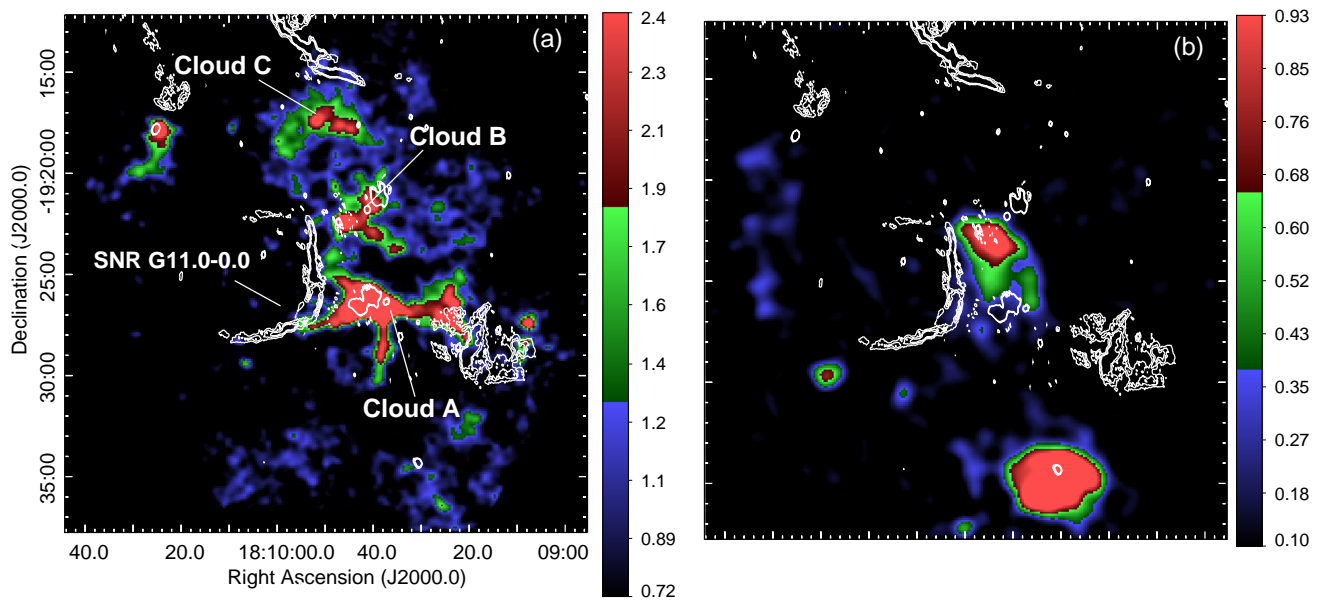}
\caption{Distribution of the $^{12}$CO J=$3-2$ gas in the field containing the 
SNR~G11.0$-$0.0 and the TeV source HESS~J1809$-$193.
\bf a) \rm Averaged intensity distribution of the $^{12}$CO in the velocity range from 16 to 
27~km~s$^{-1}$. \bf b) \rm Averaged intensity map of the molecular gas for velocities between 71 and 
78~km~s$^{-1}$. In both panels the white contours  delineate the radio continuum emission at 1.4~GHz. 
The wedge to the right of each panel shows the $^{12}$CO intensities in K~km~s$^{-1}$.}
\label{CO}
\end{figure*}

After a careful inspection of the whole $^{12}$CO~(J=$3-2$) data cube in the environment of 
HESS~J1809$-$193, we identified molecular material at different velocities that shows tight
spatial correlation with both the SNR~G11.0$-$0.0 and the $\gamma$-ray peak. The observed 
molecular clouds are limited to the velocity intervals (+16~km~s$^{-1}$, +27~km~s$^{-1}$) and (+71~km~s$^{-1}$, +78~km~s$^{-1}$). Figure~\ref{CO}(a) shows the distribution of the  $^{12}$CO emission integrated 
in the lower velocity range overlaid with contours delineating the 
location of the SNR G11.0$-$0.0. At least three clouds, which we  refer to
in this work as A, B, and C for convenience, can be distinguished
in the large-scale distribution of the molecular gas. These components seen in projection on the plane of the sky overlap the western
outer border of G11.0$-$0.0 in all its extension. Feature A overlaps the flattened and intense region of the 
radio shell towards the south and delineates the western border of the SNR. Cloud B coincides with the brightest region of HESS~J1809$-$193 \citep[depicted as a 5$\sigma$ significance contour in Fig.~1 of][]{komin08} and one 
of its extremes is in contact with the northwestern rim of G11.0$-$0.0.
Finally, the molecular component C, while located in the northern extreme of the remnant, still 
lies within the extension of the $\gamma$-ray emission. 
We adopt $\sim$21~km~s$^{-1}$ as the systemic velocity of the molecular material;  because of the so-called
kinematic distance ambiguity in the inner Galaxy, this results in a near and far kinematic
distances of $\sim$2.9 and $\sim$13.6~kpc, respectively.
Figure~\ref{CO}(b) shows the intensity of the $^{12}$CO integrated in the velocity range of 71-78~km~s$^{-1}$, superimposed with some contours of the radio continuum emission from
G11.0$-$0.0. From this image, a good correspondence can be seen between
the flat western boundary of the remnant and the molecular gas.
Interestingly, part of the cloud projects onto the brightest $\gamma$-ray emission.
For this cloud we adopt  a systemic velocity of $\sim$75~km~s$^{-1}$, which corresponds to near/far distances of 
$\sim$6/11~kpc.

\begin{figure}[h]
\centering
\includegraphics[scale=0.7]{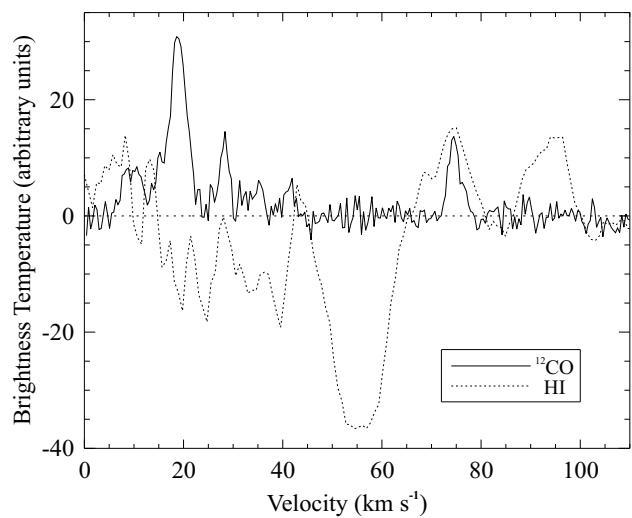}
\caption{$^{12}$CO and HI~21~cm spectra towards the molecular material identified in the field of SNR~G11.0$-$0.0. The peak brightness centered at $\sim$21~km~s$^{-1}$ corresponds to the nearby clouds
located at $\sim$3~kpc, while the peak at $\sim$75km~s$^{-1}$ is caused by background $^{12}$CO gas located at the far kinematic distance $\sim$11~kpc in the line of sight. }
\label{12CO-HI}
\end{figure}

Following \citet{roman09} we resolved the kinematic distance ambiguity for each cloud 
using the $^{12}$CO information in conjunction with the cold atomic hydrogen gas embedded within the molecular cloud. Because HI is present throughout the Galaxy, if the cloud is located at a near distance, a correlation between the $^{12}$CO emission line from the cloud and an HI self-absorption feature
should be noticed in the spectrum because the atomic gas in the interior of the cloud is colder than the rest of the Galactic atomic hydrogen located behind  the cloud along the line of sight. 
On the contrary, no absorption feature in the HI spectrum should be present for a cloud placed at the far 
kinematic distance because there is no warm Galactic HI at the velocity of the molecular cloud to be absorbed.
From the rotation curve it is evident that the warm HI that could be located beyond the far molecular cloud
can only  be associated with velocities completely different from the velocity of the cloud of interest. 
In Fig.~\ref{12CO-HI} we show the HI~21~cm and $^{12}$CO spectra in the direction of
the molecular clouds identified at $\sim$21~km~s$^{-1}$ and $\sim$75~km~s$^{-1}$. 
At the lower velocity, the HI~21~cm spectrum exhibits a deep self-absorption throughout the molecular emitting region and hence we assign to this cloud the near distance (d$\sim$3~kpc).
Conversely, the emission feature in the HI~21~cm spectrum centered at 75~km~s$^{-1}$, which is in coincidence
with a peak in the $^{12}$CO emission, revealed that this cloud is located at the far kinematic distance 
(d$\sim$11~kpc).

We estimate the mass of the structures A, B, and C observed at $\sim$21~km~s$^{-1}$ and the cloud 
identified at $\sim$75~km~s$^{-1}$. To do this, we first calculate the H$_2$ column density using the
empirical relation $N(\mathrm{H_2})=X_{\mathrm{CO}} \times W(^{12}\mathrm{CO})$, where 
$W(^{12}\mathrm{CO})$ is the integrated line CO flux and X$_{CO}$ 
is the CO to H$_2$ conversion factor for the $^{12}$CO J=$3-2$ transition. 
By assuming  that the gas is in local thermodynamic equilibrium,  
it means that all molecular levels are expected to be equally populated, and hence
we can use $X_{\mathrm{CO}}=(2\pm0.6)\times10^{20}$~cm$^{-2}$~K~km~s$^{-1}$, which is the canonical conversion factor between the H$_2$ and the
$^{12}$CO in the J=$1-0$ transition \citep{bolatto13}. 
The mass for each cloud is then determined
through the relation $M = \mu \, m_{\mathrm{H}}\, \Sigma[d^{2} \, \Omega \, N(\mathrm{H_{2}})]$, with  
$\mu$ being the mean molecular mass equal to 2.8 assuming a relative helium abundance of 25\%, 
$m_{\mathrm{H}}$ the mass of the hydrogen atom, $\Omega$ the angular
area subtended by the source, and $d$ the distance to the cloud. 
In order to estimate the total ambient proton distribution, 
we extend our analysis of the ISM properties by exploring the contribution of the atomic gas 
to the mass and density of the clouds defined by the velocity ranges used for the $^{12}$CO data. 
Under the assumption that the neutral atomic gas is optically thin, the HI column density is
given as $N(\mathrm{HI})=1.823\,\times\,10^{18}\,\int{T_{\mathrm{b}}\, dv}$ \citep{dickey90}, where
$T_{\mathrm{b}}$[K] and $v$[km~s$^{-1}$] represent the intensity and velocity of the neutral gas, respectively. 
The volume densities were estimated assuming elliptic representations of each molecular cloud 
with a depth comparable to the size of the minor axis. The individual estimates for the mass and density of all the clouds are summarized in Table~\ref{masas}.
The error in the cloud masses are estimated to be on the order of 45\%, which mainly arises from
the uncertainty in distance measurements ($\sim$30\%).

\begin{table*}
\caption{Derived parameters for the molecular clouds}
\renewcommand{\arraystretch}{1.0}                                                                                
\begin{center}
\begin{tabular}{lccccc} \hline\hline \\
Feature & Central Position &$N_{\mathrm{H_{2}}}$ & $N_{\mathrm{H}}$ & Mass       &         Total proton density  \\
& R.A., decl. (J2000.0)& $\times$10$^{21}$~[cm$^{-2}$] & $\times$10$^{21}$~[cm$^{-2}$] &   
$\times$10$^{3}$~[$M_\odot$]  & $\times$10$^{3}$~[cm$^{-3}$] \\
\hline
Cloud A at 21~km~s$^{-1}$ & $18^{\mathrm{h}}09^{\mathrm{m}}46.3^{\mathrm{s}}$,$-19^{\circ}26^{\prime}53^{\prime\prime}$&  5.3 &  2.2 & 0.8 & 1.9  \\
Cloud B at 21~km~s$^{-1}$  & $18^{\mathrm{h}}09^{\mathrm{m}}41.8^{\mathrm{s}}$,$-19^{\circ}21^{\prime}38^{\prime\prime}$ &  9.6 & 2.1 & 1.3 & 3  \\
Cloud C at 21~km~s$^{-1}$  &$18^{\mathrm{h}}09^{\mathrm{m}}51.3^{\mathrm{s}}$,$-19^{\circ}17^{\prime}45^{\prime\prime}$ &  6.4 & 2.5 & 0.7 & 2.5 \\
Cloud at 75 km~s$^{-1}$    & $18^{\mathrm{h}}09^{\mathrm{m}}41.5^{\mathrm{s}}$,$-19^{\circ}24^{\prime}33^{\prime\prime}$ &  1.5 & 0.3  & 4.2 & 0.1 \\ \hline           
\end{tabular}
\end{center}
\label{masas}
\end{table*} 

\begin{figure}[h!]
\centering
\includegraphics[scale=1]{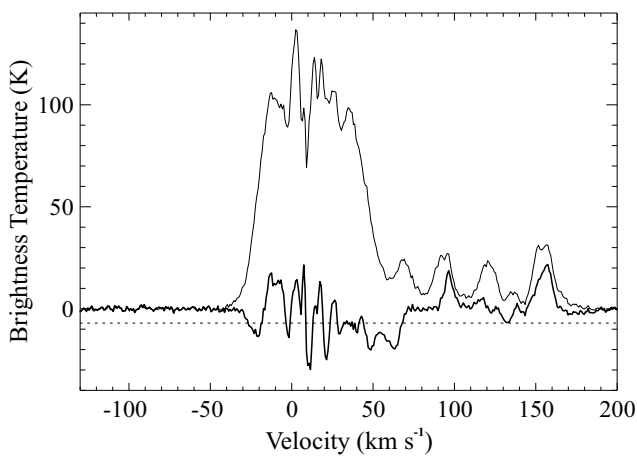}
\caption{HI emission (thin line) and absorption (thick line) spectra towards SNR~G11.0$-$0.0. The dashed line indicates absorption at the 5$\sigma$ level, where $\sigma$ is the noise level in the HI data cube.}
\label{co-hi}
\label{HIspectrum}
\end{figure}

Before discussing the possible role of the identified molecular material as 
target ISM protons to produce the observed $\gamma$ rays, we estimate the distance to SNR~G11.0$-$0.0
using the HI absorption technique. While G11.0$-$0.0 itself is relatively
weak and has a thin shell, we construct the emission spectrum against the southern part of the SNR shell 
where there is a dominant bright feature. The absorption profile was obtained by subtracting the bright 
radio shell emission from nearby regions devoid of continuum emission. Both spectra are shown in 
Fig.~\ref{HIspectrum};  the most positive-velocity reliable absorption feature is seen at 
$\sim$20~km~s$^{-1}$. Regarding the lower level absorption dip 
in the velocity range $\sim$50-65~km~s$^{-1}$, we believe that in view of its extent
and its association with  relative low-amplitude emission, this
feature may represent fluctuations in the HI emission and hence we are not confident that the absorption is 
real. Additionally, no absorption is observed at negative velocities. Therefore, we consider 
the value 20~km~s$^{-1}$ as a lower limit to the systemic velocity of G11.0$-$0.0, 
corresponding to  near and far kinematic distances of $\sim$3 and $\sim$13.8~kpc, respectively. 
However, if we regard  the feature at 50-65~km~s$^{-1}$ as reliable, the absence of absorption between these velocities and the tangent point at $\sim$167~km~s$^{-1}$ indicates that an upper limit on the distance to the SNR of $\sim$5.2~kpc can be adopted. 
The distance of $\sim$17~kpc as derived by \citet{Brogan04} using the 
$\Sigma$-D correlation between the
linear diameter of the G11.0$-$0.0 shell and its radio surface brightness is not consistent with the range of possible values 
(3.0-5.2~kpc)  that we have determined for SNR~G11.0$-$0.0 through HI absorption. 
Such a disagreement is not surprising, 
since it has long been known that the $\Sigma$-D relationship  has a large intrinsic dispersion.

Taking into account the good morphological correspondence between the remnant G11.0$-$0.0 and the 
molecular material located at the same systematic velocity, we therefore argue a physical 
connection between them. This allows us to resolve the ambiguity of the distance to G11.0$-$0.0: 
 it is located at about 3~kpc.

The question  to be addressed, therefore,  is whether the cloud components traced by the $^{12}$CO emission 
at $\sim$21~km~s$^{-1}$ can be considered a plausible astronomical environment for producing
the detected photons in HESS~J1809$-$193. To pursue this point we first calculate the $\gamma$-ray flux from HESS~J1809$-$193 taking into account that the spectrum of this source is  fit well by a  pure power law
(dF$_{\gamma}(\mathrm{E})$/dE=N$_{0}\,(\mathrm{E}/1~\mathrm{TeV})^{-\Gamma}$)
with a photon index $\Gamma$=2.23 and 
$N_{0}$=6.4$\times 10^{-12}$~TeV$^{-1}$~cm$^{-2}$~s$^{-1}$ \citep{renaud2008}, 
which implies F$_{\gamma}(>1\mathrm{TeV})= 5.2\times 10^{-12}$~ph~cm$^{-2}$~s~$^{-1}$. 
Then, we quantify the ambient density required to power the observed $\gamma$-ray flux 
through the  Eq.~(16) of \citet{Torres03}, 
given our estimate of the distance to the molecular material (d$\sim$3~kpc) 
along with other assumptions such as an efficiency $\theta$=3\% 
to convert the accelerated hadrons into cosmic rays and a supernova energy output of 10$^{51}$~erg. 
We find that at a distance of 3~kpc an ambient density of 
$\eta_{3\,\mathrm{kpc}}$$\sim$40~cm$^{-3}$ it would be necessary to produce the $\gamma$-ray emission detected above 1~TeV.   
By comparing this density with that measured in the molecular 
material comprising  Clouds A, B, and C, we conclude that it is 
dense  enough to generate the VHE emission hadronically.
We also infer an expected volume density $\eta_{11\,\mathrm{kpc}}$$\sim$470~cm$^{-3}$ for the
molecular material identified at $\sim$75~km~s$^{-1}$, which favors our interpretation that this background
gas at d$\sim$11~kpc is not related to the production of $\gamma$-rays photons in HESS~J1809-193.

\section{Summary}
At present, the most plausible origin for the $\gamma$-ray emission from HESS~J1809$-$193 has
been explained by a PWN powered by the pulsar PSR~J1809$-$1917; however,
there is no  conclusive observational evidence to support the proposed leptonic scenario. 
This  motivated us to perform the first high-sensitivity radio study of a large region
containing the TeV source at 1.4~GHz using the JVLA. In addition, we analyzed the ISM 
of this area. Although our radio observations surpass  the sensitivity of available radio images by more than an order of magnitude (noise level of our image $\sim$$0.17~$mJy~beam$^{-1}$), 
there is no trace of radio emission associated with the X-ray PWN driven by PSR~J1809$-$1917 nor an indication of the existence of a host SNR. We also analyzed the $^{12}$CO in the vicinity of 
HESS~J1809$-$193 to investigate
the possibility of any other candidate in the field of view that could explain
the production mechanism of the high-energy $\gamma$-rays. 
As a result of this analysis, we discovered a system of molecular clouds (centered at v$\sim$21~km~s$^{-1}$) that includes three main components (called in this work clouds A, B, and C).
This material is well correlated with the TeV source and matches the morphology of the western edge of the SNR~G11.0$-$0.0. 
The distance to these clouds was constrained to be $\sim$3~kpc based on the association between $^{12}$CO emission and HI-self absorption features, which also coincides within uncertainties with the kinematical distance  
derived for G11.0$-$0.0. We calculated the total mass and total
proton density of the A, B, and C clouds (taking into account contributions from both HI and $^{12}$CO,
the values are 
$M\simeq 3 \times$10$^{3}$~$M_{\odot}$ and $\eta \simeq 7.4 \times$10$^{3}$~cm$^{-3}$, see Table~\ref{masas}),
which we assumed are interacting with the remnant
G11.0$-$0.0, and found that they indeed satisfy the required amount of target material 
to explain the observed TeV $\gamma$-ray flux from HESS~J1809$-$193 as produced by hadrons.

We finally note that our analysis does not rule out a hybrid
scenario in which contributions from relativistic leptons via inverse Compton process 
in the PSR~J1809$-$1971 operate together with the hadronic process that we have proposed in
this work. In addition, star formation activity can play a role in HESS~J1809$-$193 as several
bright HII regions lie within the boundary of the VHE source.    

\begin{acknowledgements}

This research was
partially funded by Argentina Grants awarded by
 ANPCYT: PICT 0571/11 and 0902/13; CONICET: 0736/11; and University of Buenos Aires (UBACYT). G.C., E.G., and A.P. are Members of the Carrera del Investigador Cient\'\i fico of CONICET, Argentina.

\end{acknowledgements}

\bibliographystyle{aa}  
\bibliography{astroph-castelletti.bib}
\IfFileExists{\jobname.bbl}{}
{\typeout{}
\typeout{****************************************************}
\typeout{****************************************************}
\typeout{** Please run "bibtex \jobname" to optain}
\typeout{** the bibliography and then re-run LaTeX}
\typeout{** twice to fix the references!}
\typeout{****************************************************}
\typeout{****************************************************}
\typeout{}
}

\end{document}